\documentclass{aa501}
\usepackage{graphicx,natbib,amssymb}

\newcommand{\ls}
 {\mathrel{\hbox{\rlap{\hbox{\lower4pt\hbox{$\sim$}}}\hbox{$<$}}}}
\newcommand{\gs}
 {\mathrel{\hbox{\rlap{\hbox{\lower4pt\hbox{$\sim$}}}\hbox{$>$}}}}
\newcommand{\degg}{\hbox{$^\circ$}}

\newcommand{\arcs}{\hbox{$^{\prime\prime}$}}
\newcommand{\et}{et al.\ }

\newcommand{\asca}{{\it ASCA}}

\newcommand{\sax}{{\it Beppo-SAX}}
\newcommand{\xmm}{{\it XMM-Newton}}

\def\la{\mathrel{\hbox{\rlap{\hbox{\lower4pt\hbox{$\sim$}}}{\raise2pt\hbox{$
<$}}
}}}
\def\ga{\mathrel{\hbox{\rlap{\hbox{\lower4pt\hbox{$\sim$}}}{\raise2pt\hbox{$
>$}}
}}}

\begin{document}

\title{Soft X-ray emission lines in the afterglow spectrum of GRB
011211:- a detailed XMM-Newton analysis}
\titlerunning{Soft X-ray emission lines in GRB 011211}

\author
{J.N. Reeves\inst{1}, D. Watson\inst{1}, J.P. Osborne\inst{1},
K.A. Pounds\inst{1}, P.T. O'Brien\inst{1}}

\institute{X-ray and Observational Astronomy Group, 
Department of Physics and Astronomy, University of Leicester,
University Road, Leicester LE1 7RH; U.K.}

\authorrunning{J. Reeves et al.}

\date{Submitted June 2002}

\abstract {We report on an \xmm\ observation of the X-ray
afterglow of the Gamma Ray Burst GRB 011211, originally detected by
\sax\ on 11th December 2001. The early afterglow
spectrum obtained by \xmm, observed 11 hours after the initial burst,
appeared to reveal decaying H-like K$\alpha$ emission
lines of Mg, Si, S, Ar and Ca, arising in
enriched material with an outflow velocity of order 0.1c (Reeves \et 2002).
This was attributed to matter ejected from a massive stellar progenitor
occurring shortly before the burst itself.
Here, we present a detailed
re-analysis of the \xmm\ EPIC observations of GRB 011211. In
particular, we show that the detection of the soft X-ray line emission
appears robust, regardless of detector background, calibration, spectral
binning, or the spectral model that is assumed. We
demonstrate that thermal emission, from an optically thin plasma, is
the most plausible model that can account for the soft X-ray emission,
which appears to be the case for at least two burst afterglow
spectra observed by \xmm. 
The X-ray spectrum of GRB 011211 evolves
with time over the first 12 ksec of the \xmm\ observation, 
the observations suggest that thermal emission dominates the early
afterglow spectrum, whilst a power-law component dominates the
latter stages. Finally we estimate the mass of the ejected material
in GRB 011211 to be of the order 4-20 solar masses.}

\maketitle


\section{Introduction}

The study of X-ray afterglow emission is crucial to understanding
the nature of gamma ray bursts and their progenitors, as
X-ray spectroscopy can reveal details of the environment of the burst
explosion. For instance, observations of some X-ray afterglows
with \sax, \asca, and {\it Chandra} 
have revealed strong iron K$\alpha$ emission lines
(e.g. GRB 000214, Antonelli \et 2000;
GRB 991216, Piro \et 2000; GRB 970828, Yoshida \et 1999; GRB 970508, Piro
\et 1998), with line equivalent widths of up to 
several keV; typically the lines are detected at 
$\sim3\sigma$ confidence. An iron K-shell absorption feature
was also reported in the
prompt \sax\ spectrum of one burst, GRB 990705 (Amati \et 2000,
Lazzati \et 2001).
The line observations appear to support models where 
the burst explodes into an enriched, high density medium, favoring a
massive stellar progenitor for long duration bursts.
In many cases (e.g. Piro \et 2000,
Antonelli \et 2000), large masses of iron are required to account for
the line emission. One possible source of the iron is in a shell of
distant (R~$\sim10^{16}$~cm) material ejected from a supernova (the
`supranova' model, Vietri \& Stella 1998)
which would have to occur at least several months {\it prior} to the burst
event, in order to allow sufficient time for large quantities of iron
to form (e.g. Vietri \& Stella 1998). Alternatively, the
iron K emission could arise through the interaction of a magnetically
driven wind with the envelope of the massive progenitor star (Rees \&
Meszaros
2000); here the distances involved are much smaller
(R$\sim10^{13}$~cm) and consequently the high masses of iron
and long time delays between supernova and burst are not required.

The high throughput of \xmm, compared with \sax\ or {\it Chandra}, makes it
the best available telescope with which to constrain X-ray line
emission in GRBs,
potentially providing a powerful diagnostic
for discriminating between models of the burst progenitor.
In an analysis of the \xmm\ observation of the afterglow of
gamma ray burst GRB 011211, Reeves \et (2002) reported the first
detection of emission lines other than iron in a GRB afterglow spectrum;
specifically the decaying line emission from the hydrogenic states of
Mg, Si, S, Ar and Ca.
Furthermore the energies of the X-ray lines appeared to be
offset from the known redshift for the host galaxy of GRB 011211 
($z=2.140\pm0.001$,
Holland \et 2002), implying the line emitting matter was outflowing with
a velocity of $\sim$~0.1c from the site of the GRB progenitor.
This result was interpreted as possible evidence of a supernova
explosion occurring within days of the burst itself,
with the X-ray line emission arising from
matter in the expanding supernova shell.

Recently, Borozdin \& Trudolyubov (2002) have claimed that the detection
of soft X-ray emission lines from light metals in GRB 011211 may
not be robust, as the observations could be contaminated by the
background
of the \xmm\ EPIC-pn detector. In addition, Rutledge \& Sako (2002)
perform simulations which suggest that the significance of
the emission features may be lower, at the level (at best) of
$\sim98.8\%$ confidence.
In this paper, we describe our 
analysis of the \xmm\ data on GRB 011211 in detail. We confirm
that the soft X-ray emission features in GRB 011211 are
robust, the set of lines being detected to a good level of confidence
(at $>99$\% significance in all our tests). 
Both the F-test and Monte-Carlo simulations
are used to determine the statistical significance of the line 
emission features, making no prior assumption about the rest energies
or redshift of the line emitting material.  
We also demonstrate that the results do not
depend on the background subtraction of the EPIC-pn detector,
instrument calibration, or the spectral binning used
and show that the result is independent of the spectral model
that is assumed.
In section 2 we review the properties of the burst, the
\xmm\ observations and basic data analysis and
in section 3 detail the spectral analysis of the afterglow during the
first 5 ksec when the lines appear most prominent.
The time dependent spectral properties of GRB 011211 are also outlined,
whilst a likely mechanism of the line emission is discussed
and a mass for the ejected material in GRB 011211 is derived.
Throughout this paper
we adopt a cosmology of $H_{0}= 75$~km~s$^{-1}$~Mpc$^{-1}$
and $q_{0}=0.1$. Unless otherwise stated, all errors are quoted at
68\% confidence (i.e. $\Delta\chi^2=1.0$ or 2.3 for 1 or 2 interesting
parameters respectively).

\section{XMM-Newton Observations of GRB 011211}

The Gamma Ray Burst GRB 011211 was first detected on December 11th
2001 at 19:09:21 (UT), by Beppo-SAX (Frontera \et 2001).
The burst duration was 270s, making GRB 011211 the longest burst
yet observed by Beppo-SAX, with a
peak flux (40-700 keV) of $5\times10^{-8}$~erg~cm$^{-2}$~s$^{-1}$.
Spectroscopy of the transient
optical afterglow revealed several absorption lines at a
redshift of $z=2.140\pm0.001$ (Holland \et 2002).
Optical imaging (Burud \et 2001, Fox 2002) has also linked the optical
transient to extended emission, the probable host galaxy of
magnitude $m_{v}=25.0\pm0.5$. Assuming the absorption system arises from the
host galaxy of the GRB, then the total
equivalent isotropic luminosity for GRB 011211 is 
$5\times10^{52}$~erg.

\begin{figure}
\includegraphics[height=\columnwidth,angle=-90,clip=]{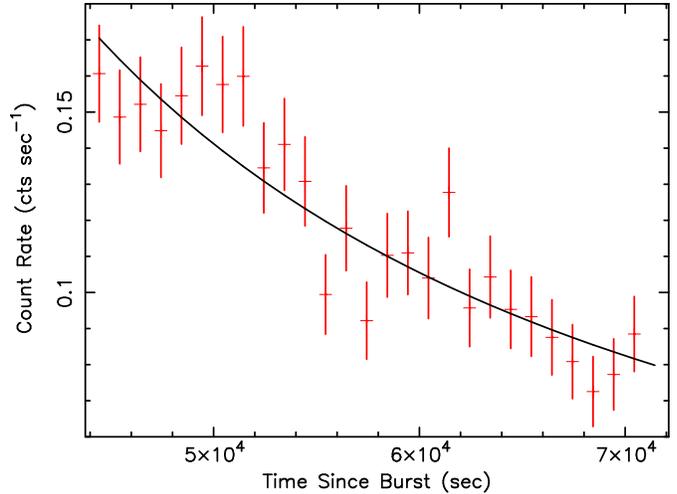}
\caption{EPIC 0.2--10\,keV lightcurve (coadded pn and MOS).  A power-law
         decay fits the data well, the best-fit power-law, (with a decay
         index of $1.6\pm0.1$) is plotted.  Time since the GRB is plotted on
         the abscissa.}
\label{lightcurve}
\end{figure}

The \xmm\ (Jansen \et 2001) observations of
GRB 011211 started at 06:16:56 (UT)
on December 12th 2001, $\sim11$ hours after the initial burst (Santos-Lleo
\et 2001). Data from
EPIC (European Photon Imaging Cameras) have been analysed, using
both the MOS CCDs (Turner \et 2001) with the medium filter and
pn CCDs (Struder \et 2002) using the thin filter.
The total observation duration (for EPIC-pn) was 27 ksec.
Approximately 4.2 ksec into
the observation, the spacecraft pointing was moved from its
original position at co-ordinates (2000) 11h 15\arcmin\ 16.4\arcs, 
--21\degg 55\arcmin\ 44.8\arcs, to 
new boresight co-ordinates at 11h 15\arcmin\ 18.0\arcs, 
--21\degg 56\arcmin\ 53.0\arcs,  
in order to place the source away from the EPIC-pn chip gaps. The
time-averaged 0.2--10 keV flux was
$1.7\times10^{-13}$~erg~cm$^{-2}$~s$^{-1}$ (corresponding to an 0.6--30 keV,
rest frame afterglow luminosity of $7\times10^{45}$~erg~s$^{-1}$),
decreasing during the observation according to a power-law, with decay rate
$F(t)\propto t^{-\alpha}$, $\alpha = 1.6\pm0.1$ (Fig.~\ref{lightcurve}).  The X-ray decay is
notably steeper than reported for the same period in the optical ($\alpha =
0.83\pm0.04$, Holland et al.\ 2002).

\begin{figure}
\includegraphics[width=\columnwidth]{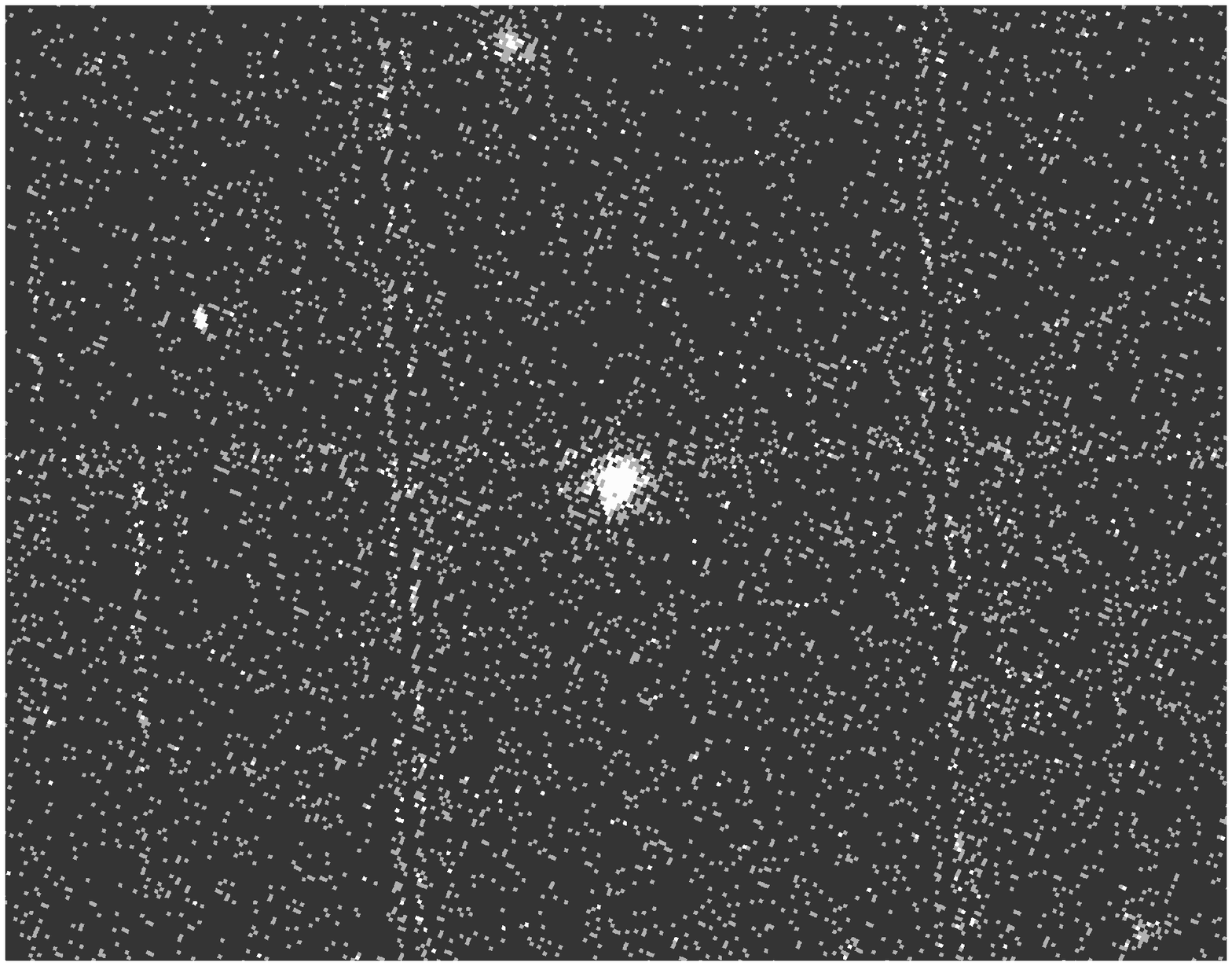}
\includegraphics[width=\columnwidth]{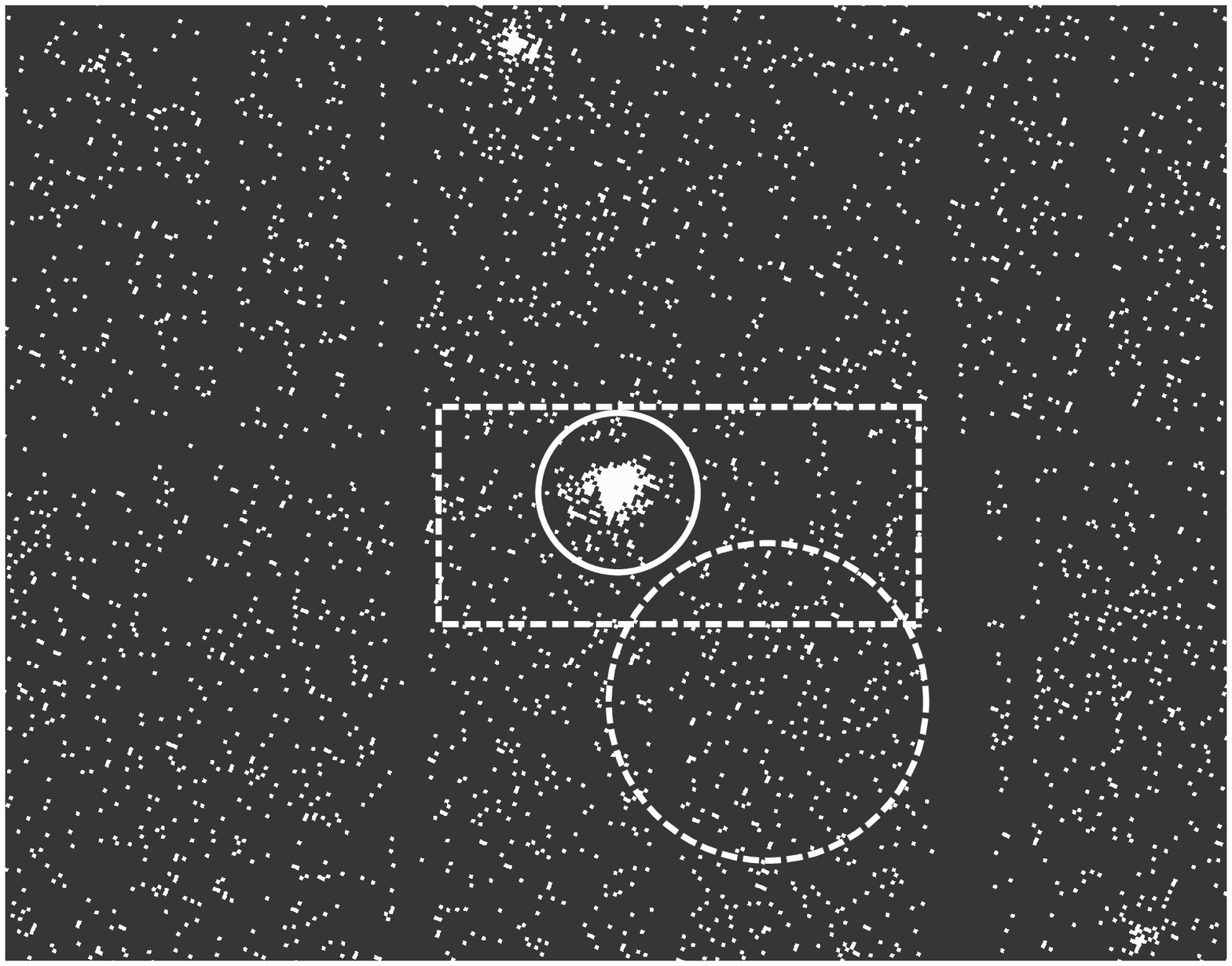}
\caption{XMM-Newton images of the field around GRB 011211, during the 
first 5 ksec of observation; (a) \textit{top} shows the unscreened image, whilst
(b) \textit{bottom} shows the screened image, used for extraction of all
spectra. Here the bold circle indicates the source spectral 
extraction region used, of radius of 40\arcs. The dashed areas
indicate the background regions used in the extractions; (i) a box
along the edge of the chip (but excluding the source region) or (ii) an
offset circle of radius 80\arcs.}
\label{images}
\end{figure}

\subsection{Extraction of the XMM-Newton data}

We initially concentrate our analysis on the first 5 ksec of
observation, 
where the prominent soft X-ray emission lines are observed. The analysis
focuses on the EPIC-pn spectrum, which due to the use of the thin 
filter, has a much higher count rate
($>6\times$) than
the EPIC-MOS, having first checked that the MOS and pn data are
self-consistent. All data reduction was performed using the XMM-SAS
(v5.3).

Fig.~\ref{images}(a) shows the unscreened EPIC-pn field around
GRB 011211, the afterglow itself being the bright source
located close to the edge of the EPIC-pn chip
gap, but near to the centre of the image. 
The background count rate is clearly higher around the edges of the
pn chip in the unscreened dataset, as noted by 
Borozdin \& Trudolyubov (2002), who
claim that inclusion of the noisy pixels in the background spectrum
could effect the detection of soft X-ray lines. However in order to
produce a clean image and event list for GRB 011211, the dataset was
screened to include only good X-ray events (using the selection
expression `FLAG=0', in the SAS tool evselect) using only
the well-calibrated single or double
pixel events (i.e. PATTERN$<$=4). An energy cut was also applied,
excluding all data below 200 eV and above 10 keV.
The resulting image is shown in Fig.~\ref{images}(b), which shows a smooth
distribution of
background counts, regardless of position on the EPIC-pn chips.

\begin{figure}
\includegraphics[height=\columnwidth,angle=-90,clip=]{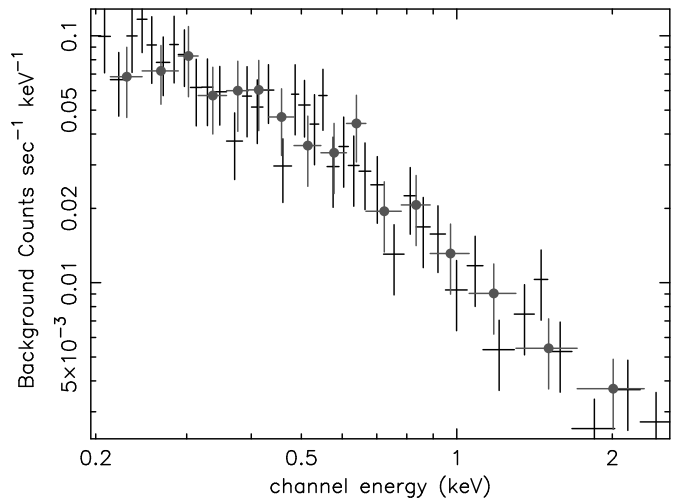}
\caption{Background spectra used for spectral analysis during the
initial 5 ksec, the grey
circles show the spectrum extracted from the offset circular background
region, whilst the crosses show the background spectrum extracted from
the box region, located close to the chip edge. The two background
spectra are identical, and demonstrate that the background spectrum 
obtained is both featureless and 
independent of the position on the pn chip. Overall the
background count rate is $\sim10\%$ of the net count rate from GRB 011211.}
\label{back_specs}
\end{figure}

Source spectra for GRB 011211 were subsequently extracted from a
circular region of 40\arcs\ radius, centred on the source,
accounting for the decrease in flux incurred through photons falling
onto the chip gaps. A background spectrum was
taken from a circular region of radius 80\arcs, using a source-free region
located close to the source (Fig.~\ref{images}b). The background spectrum
obtained is shown in Fig.~\ref{back_specs}; after applying the above
screening the background spectrum is smooth, unlike the background
spectrum obtained by Borozdin \& Trudolyubov (2002), who using
unscreened data obtain a `spike' in their background spectrum near 
0.7 keV. However,
in order to ascertain whether the choice of background region can
affect the spectrum of GRB 011211, we also use a background
extracted from a box-shaped 
area located close to the EPIC-pn chip gap (Fig.~\ref{images}b), but excluding
the source region. The background spectra obtained for both regions are
identical (Fig.~\ref{back_specs}), hence the choice of
background region does not effect the background subtracted
source spectrum of
GRB 011211. When normalized to the size of the 
source extraction region,
the background contributes $\sim10\%$ of the net source count rate.

\section{Spectral Analysis}

The background subtracted source spectrum of GRB 011211, 
from the first 5 ksec of the observation, was fitted in 
\textsc{xspec v11.1}, 
using a minimum of 12 source counts per bin. An 
absorbing column of $4.2\times10^{20}$~cm$^{-2}$, due to our own galaxy, 
was included in all the spectral fits. A simple power-law fit (model
1) gives a barely (10\% null hypothesis probability) 
acceptable fit to the data
($\chi^{2}/d.o.f=54.2/45$, where 
d.o.f is the number of degree of freedom), with a power-law photon
index of 
$\Gamma=2.18\pm0.10$. The model fits to GRB 011211 are summarized in
Table 1.

\begin{figure}
\includegraphics[height=\columnwidth,angle=-90,clip=]{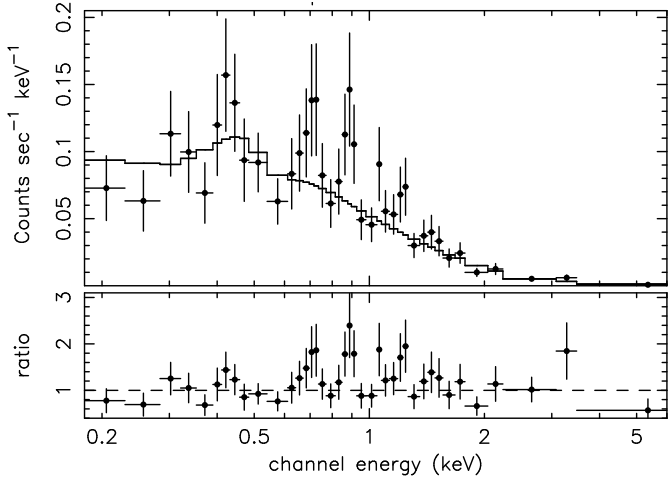}
\caption{The background subtracted  
\xmm\ EPIC-pn spectrum of GRB 011211,
from the first 5 ksec of observation. Energy (units keV), 
in the observed frame,
is plotted on the abscissa. In the upper panel,  
the EPIC-pn data points are
plotted with filled circles, whilst the solid line shows a power-law
model, folded through the EPIC-pn response function. The lower panel
shows the ratio 
of the data to the power-law fit; an excess of
counts is observed between 0.5 and 1.5 keV in the spectrum. This
excess can be well fitted with at least 3 emission lines, at
$0.71\pm0.01$~keV, $0.88\pm0.01$~keV and $1.21\pm0.02$~keV
respectively.}
\label{pow_spec}
\end{figure}

\subsection{Soft X-ray Emission Line Fits to GRB 011211}

Inspection of the power-law fit to the spectrum of GRB 011211 shows 
a clear excess present in the data residuals 
between 0.5 and 1.5 keV (Fig.~\ref{pow_spec}). The deviations are 
considerably larger than the current calibration uncertainty of
the EPIC-pn detector, which is of the order $\sim5\%$; adding a
further systematic error of 5\% to the data has no noticeable effect
on the spectral fits. Hence, following the procedure of 
Reeves \et (2002), we attempt 
to fit this excess with a set of emission lines. Initially 3
narrow spectral lines were added (model 2), allowing the energy 
and flux of each line to vary, whilst the line widths
were kept negligibly small (at $\sigma=10$~eV) 
in the fits, as the lines are unresolved compared
to the detector resolution. The best fit 
(observed frame) energies of the lines were then $0.71\pm0.02$~keV, 
$0.88\pm0.01$~keV and $1.21\pm0.02$~keV, with line fluxes of  
$(9.9\pm3.4)\times10^{-6}$~photons~cm$^{-2}$~s$^{-1}$, 
$(7.7\pm2.3)\times10^{-6}$~photons~cm$^{-2}$~s$^{-1}$ and 
$(4.0\pm1.4)\times10^{-6}$~photons~cm$^{-2}$~s$^{-1}$ respectively. The 
fit is significantly improved, at 99.7\% confidence, 
upon the addition of these lines 
($\chi^{2}/d.o.f=33.7/39$, $\Delta\chi^{2}=20.5$ for 6 extra degrees 
of freedom), when an F-test is employed (Bevington 1992). Note that
even when fitted 
individually, the lines are detected at 98\%, 97\% and 93\%
confidence (for Si, S and Ar respectively), with the addition of 2
extra degrees of freedom (energy, flux) for each line.  

\begin{figure}
\includegraphics[height=\columnwidth,angle=-90,clip=]{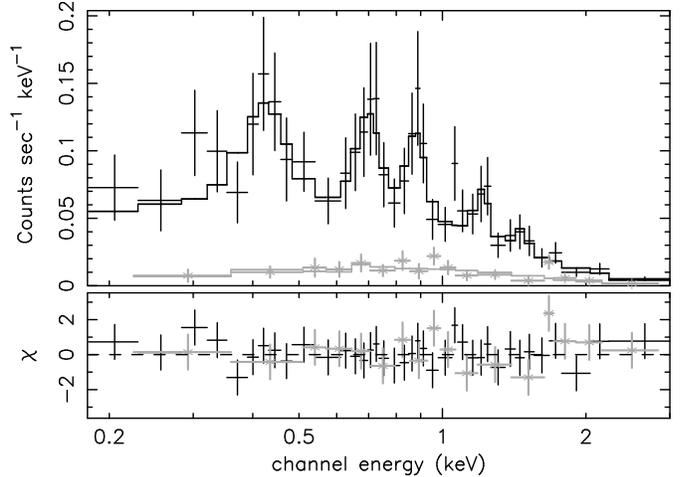}
\caption{The \xmm\ spectrum of GRB 011211, fitted with the redshifted
emission line model. The EPIC-pn spectrum is shown in black 
whilst the lower signal to noise EPIC-MOS spectrum is shown
in grey. 
The lines fitted to the data correspond to \ion{Mg}{xi} (or \ion{Mg}{xii}), \ion{Si}{xiv}, \ion{S}{xvi}, \ion{Ar}{xviii} and \ion{Ca}{xx} at a redshift of
$z=1.90\pm0.06$, implying an outflow velocity of $\sim0.1c$ for the
emitting material.}
\label{line_spec}
\end{figure}

\begin{table}
\centering
\caption{Summary of X-ray spectral fits to the first 5 ksec of
observation for GRB 011211. $^{a}$ Fit statistic, in terms of
$\chi^{2}$ divided by degrees of freedom. $^{b}$ Significance of fit
improvement, compared to a power-law, from using an F-test. $^{f}$
indicates parameter is fixed in fit.}

\begin{tabular}{lccc}
\hline\hline                 
Model & $\Gamma$ or kT & $\chi^{2}/d.o.f.$$^{a}$ & Prob$^{b}$ \\   
\hline

1. PL Only & $2.18\pm0.10$ & 54.2/45 \\

2. PL + 3 lines & $2.15\pm0.11$ & 33.7/39 & 99.7\% \\

3. PL + 5 lines & $2.10\pm0.15$ & 29.0/35 & 99.3\% \\

4. PL + z-lines & $2.12\pm0.12$ & 34.3/39 & 99.5\% \\

5. PL + VMEKAL & $\Gamma=2^{f}$ & 38.1/41 & 99.8\% \\

& $kT=4.1\pm1.6$ \\

\hline

\end{tabular}
\end{table}

Examination of the data/model residuals in Fig.~\ref{pow_spec} suggests that 
two further lines can be added to the spectrum (model 3), with observed 
energies $0.44\pm0.04$~keV, $1.46\pm0.04$~keV and fluxes
$(1.0\pm0.8)\times10^{-5}$~photons~cm$^{-2}$~s$^{-1}$ and
$(1.9\pm1.1)\times10^{-6}$~photons~cm$^{-2}$~s$^{-1}$ respectively, 
although the improvement in fit statistic is marginal 
for these last two cases ($\chi^{2}/d.o.f=29.0/35$). 
Table 2 lists the properties of each of 
the individual lines added to the X-ray spectra. 
Fig.~\ref{line_spec} shows the EPIC-pn and EPIC-MOS data fitted with the 5 emission
lines. Unfortunately none of the lines are significantly detected in
the MOS CCDs, due to the lower count rate,
although the data for both detectors are consistent 
(e.g. see the residuals in the lower panel). 

\begin{table*}
\centering
\caption{Soft X-ray line fits to GRB 011211. $^{a}$ Known (laboratory)
energy of emission line (in keV); $^{b}$ observed energy (in keV) of
emission line in spectrum; $^{c}$ measured rest-frame (at z=2.14) 
energy (keV) of the emission line; $^{d}$ redshift for the
emission line, calculated from $^{a}$ and $^{b}$. Observed 
line flux, in units $10^{-6}$~photons~cm$^{-2}$~s$^{-1}$. }

\begin{tabular}{lccccc}
\hline\hline                 

Ident & $E_{line}$$^{a}$ 
& E$_{obs}$$^{b}$ & E$_{rest}$$^{c}$ & redshift$^{d}$ &
Flux$^{e}$ \\
\hline

\ion{Mg}{xi} & 1.35 & $0.44\pm0.04$ & $1.38\pm0.10$ & $2.03\pm0.22$ &
$10\pm8$ \\

\ion{Si}{xiv} & 1.99 & $0.71\pm0.02$ & $2.22\pm0.05$ & $1.82\pm0.06$ & $9.9\pm3.4$ 
\\

\ion{S}{xvi} & 2.60 & $0.88\pm0.01$ & $2.77\pm0.05$ & $1.94\pm0.02$ & $7.7\pm2.3$ 
\\

\ion{Ar}{xviii} & 3.30 & $1.21\pm0.02$ & $3.80\pm0.06$ & $1.73\pm0.05$ & $4.0\pm1.4$
\\

\ion{Ca}{xx} & 4.07 & $1.46\pm0.04$ & $4.58\pm0.12$ & $1.79\pm0.07$ & $1.9\pm1.1$ 
\\

\hline

\end{tabular}
\end{table*}

When converted to the rest frame of the burst (at z=2.14), the energies 
of all 5 possible lines are $1.38\pm0.10$~keV, $2.22\pm0.05$~keV, 
$2.77\pm0.05$~keV, $3.80\pm0.06$~keV and $4.58\pm0.12$~keV. 
Interestingly these lines all appear 
to be offset from the known energies of the strong 
K-shell lines of \ion{Mg}{xi} (1.35 keV), 
\ion{Si}{xiv} (1.99 keV), \ion{S}{xvi} (2.60 keV), \ion{Ar}{xviii} (3.30 keV) and \ion{Ca}{xx} 
(4.07 keV). With the exception of \ion{Mg}{xi} (the most tentative detection), 
these lines appear to correspond to the 
Lyman-$\alpha$ transitions of the most abundant 
elements, and are the strongest emission lines
one would expect to observe at these X-ray energies for either a collisionally
ionised plasma (of temperature a few keV) or a highly photoionised plasma.

Having established the likely identity of the soft X-ray line 
features in GRB 011211, we then calculated the observed redshift, 
and thus the outflow velocity of the line emitting material. 
Fitting the redshift of the lines individually, one finds 
values of $z=2.03\pm0.22$ (\ion{Mg}{xi}), $z=1.82\pm0.06$ (\ion{Si}{xiv}), 
$z=1.94\pm0.05$ (\ion{S}{xvi}), $1.73\pm0.05$ (\ion{Ar}{xviii}) and $1.79\pm0.07$ 
(\ion{Ca}{xx}). Thus with the exception of \ion{Mg}{xi} (which at the detector
resolution is unresolved from \ion{Mg}{xii} and also L-shell emission
from Fe or Ni) the lines all appear to be 
significantly offset from the known redshift for GRB 011211 of 
$z=2.140\pm0.001$, corresponding to an outflow velocity of 
$\sim0.1c$. In order to test the significance of this, we fixed the lines at
these known rest-frame energies, 
but allowed a single redshift (between $z=0$ and $z=10$, in steps
$\Delta z=0.05$) for the entire line set to vary
(model 4). The redshift calculated for the line
set is $z=1.90\pm0.06$, corresponding to an outflow velocity of
$0.085\pm0.02c$ or $\sim25000$~km~s$^{-1}$ from the GRB progenitor. 
Overall, the fit statistic
improves to $\chi^{2}/d.o.f=34.3/39$. When compared to the
power-law fit (for 6 extra degrees of freedom), 
the addition of the redshifted line set  
is significant at 99.5\% confidence from an F-test. 

\begin{figure}
\includegraphics[height=\columnwidth,angle=-90,clip=]{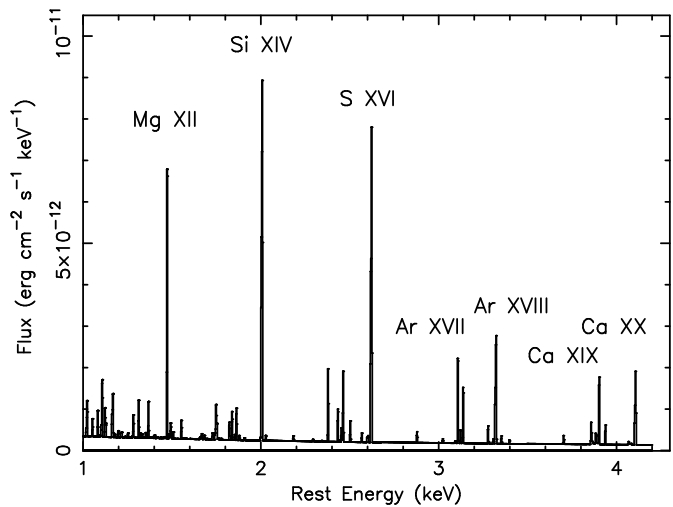}
\caption{Rest-frame thermal emission model showing the relative
strengths of the 
major emission lines (and continuum) for a plasma of temperature 4 keV.}
\label{mekal}
\end{figure}

\subsection{Thermal Emission and Ionised Reflection models}

Finally, we attempted to fit the data with emission from a 
collisionally ionised optically thin plasma, with variable elemental 
abundances (hereafter the `thermal'
emission model), using the \textsc{vmekal}
code within \textsc{xspec} (Mewe \et 1985). Initially, 
a thermal emission component of temperature $kT=4.1\pm1.6$~keV 
was added {\it in addition} to an
underlying power-law component (model 5). The improvement in fit obtained 
($\Delta\chi^{2}/d.o.f=38.1/41$) is significant when compared to
the pure power-law fit (model 1); the thermal emission component
is required at 99.8\% confidence from an
F-test, for four extra degrees of freedom (temperature,
abundance, redshift and luminosity).  
Note that the spectrum during the initial 5~ksec 
can be equally well fitted with pure
thermal emission (i.e. the power-law component is not statistically
required) the fit statistic is then $\chi^{2}/d.o.f=38.5/43$. The
rest-frame model spectrum using the \textsc{vmekal} component is shown in 
Fig.~\ref{mekal}. It can be seen that for these elements, the hydrogenic states
dominate (e.g. \ion{Si}{xiv} at 2.0 keV), although there is also some
contribution from the He-like species as well (e.g. \ion{S}{xv}, \ion{Ar}{xvii}, \ion{Ca}{xix}). 

\begin{figure}
\begin{center}
\includegraphics[height=\columnwidth,angle=-90,clip=]{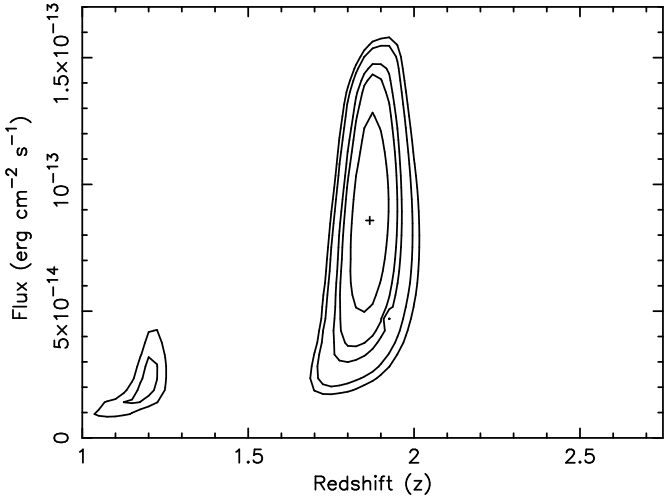}
\end{center}
\caption{Confidence contour plot of measured flux  
against redshift for the thermal line emission model,
added {\it in addition} to a power-law component. The contours represent
the 68\%, 90\%, 95\%, 99\% confidence levels to the
spectral fit, whilst the redshift has been varied (in $\Delta z=0.05$ 
increments) between $z=1$ and $z=3$. The thermal emission is
required in the data at $>99\%$ confidence, whilst the need for
an outflow velocity, with respect to the GRB rest frame, is also
required at $>99\%$ confidence.}
\label{contours}
\end{figure}

We also used this model 
to provide an independent determination of the redshift of the 
line emitting material, where no assumption has been made about the
likely identity of the lines; the best-fit redshift 
was $1.89\pm0.08$, in excellent agreement with the simple line fit
derived above. A contour plot showing the 68\%, 90\%, 95\% and 99\% 
confidence levels for the line emitting material is shown in Fig.~\ref{contours},
a redshift of $z=2.14$ can be ruled out at $>99\%$
confidence. Rutledge \& Sako (2002) 
suggest that emission at z=1.2 and z=2.75 are also plausible
solutions to the data, clearly these are ruled out from our fits,
although there is a local minimum at $z\sim1.2$ (which can be
excluded at $\sim99\%$ confidence). Nonetheless, 
we consider both the lower redshift of
$z=1.2$ and the high redshift value of $z=2.75$ to be physically
implausible, as this would require either an extreme blueshift
or for the GRB to reside at a higher redshift than the
measured value (from the optical spectrum) of $z=2.14$. 

The relative abundance of the light metals (Mg, Si, S, Ar, Ca) 
was found to be 
$\sim9\times$ the solar value, with a 90\% {\it lower-limit} of
$4\times$~solar. Interestingly, unlike the other elements,
an over abundance of iron is not required, the 90\% upper limit for the
iron abundance was found to be $<1.4\times$~solar, consistent with the
lack of detection (EW$<400$~eV, rest frame) 
of a strong iron K-shell line in the data. 

We also attempted to fit the spectrum with a reflection model (Ballantyne
\& Ramirez-Ruiz 2001), where the X-rays scatter off the walls of a funnel
excavated in the envelope of a stellar progenitor (the so-called
``nearby reprocessor'' model, Rees \& Meszaros 2000). Firstly, we
tried a pure reflection component (i.e. just the 
scattered X-ray emission). Allowing the
ionisation state of the reflecting material to vary results in an
ionisation parameter log~$\xi>4$ (with $\Gamma=2.0$) and a fit
statistic of $\chi^{2}/d.o.f=55.3/44$. In this
fit, the spectrum resembles a simple power-law, as all the elements 
are fully ionised, and the reflector is a perfect
mirror. If one forces the ionisation parameter to be log~$\xi<4$, then
the fit statistic is worse ($\chi^{2}/d.o.f=66.7/43$), whilst the
incident power-law becomes steep ($\Gamma\sim3$). The
poor-fit results from the strong iron line or edge present in the
model, and from the shape of the continuum, 
which is {\it concave} in the reflection model (but is {\it convex} in the
actual data). Allowing a
direct power-law component in addition to the reflector does not
improve the fit further. Hence it is difficult to reconcile the current
reflection models with the spectrum of GRB 011211, although we note that at
present, this model does not include emission from S, Ar
or Ca. 

\subsection{Monte-Carlo Simulations}

We have used Monte-Carlo techniques 
to provide an independent estimate of the significance 
of the line emission fits to GRB 011211. This is
important given the low photon number statistics of the
data, where the distribution of counts may become non-Gaussian
(Protassov \et 2002). 

In order to test the significance of the line emission, 
10000 random spectra from the EPIC-pn detector were generated, 
using a simple power-law
emission model (including absorption from our own Galaxy), with a
photon index of $\Gamma=2.2$. An exposure time of 5
ksec (accounting for detector deadtime) 
and an X-ray flux identical to that of GRB 011211 
($F_{0.2-10keV}=2.1\times10^{-13}$~erg~cm$^{-2}$~s$^{-1}$) were used,
as well as a background level $\sim10$\% of the source count rate, 
so that the signal to noise of the simulated spectra were
comparable to the actual source spectrum for GRB 011211. The simulated
spectra were rebinned to obtain a minimum of 12 source counts in each
spectral channel. 

We proceeded to fit the simulated data, 
using (i) a pure power-law continuum and (ii) a model consisting of 
a power-law plus three emission lines,
corresponding to the strongest lines (Si, S, Ar)
observed in the actual GRB 011211 spectrum. 
The difference in $\Delta\chi^{2}$ between
the power-law and line emission models was noted for each simulated
spectrum, in order to test whether the apparent line
emission could be produced purely through random, Poisson
noise. Intially, the line energies were fixed at the rest-frame
values corresponding to the K$\alpha$ transistions of \ion{Si}{xiv}, \ion{S}{xvi}
and \ion{Ar}{xviii}, but note that the redshift was allowed to vary between 
$z=0$ and $z=10$ (stepsize $\Delta z=0.05$) when fitting the trial spectra.
{\it Hence no prior assumption is made about the 
redshift of the line emitting material.} 
In addition, the line flux for each line was allowed to
vary in the fits. 
Upon fitting the simulations we found that only 5
in 10000 of the randomly generated power-law spectra could yield a similar
decrease in chi-squared
($\Delta\chi^{2}\sim15$), when compared to the real data,
after the addition of the three emission lines. The likelihood (or
null hypothesis probability) 
that the lines could arise purely from Poisson noise in this test 
is $<0.05\%$; hence the simulations indicate that the soft X-ray line
emission is detected at $99.95\%$ confidence, in agreement with the results
obtained through the F-test. 

\begin{figure}
\includegraphics[width=\columnwidth]{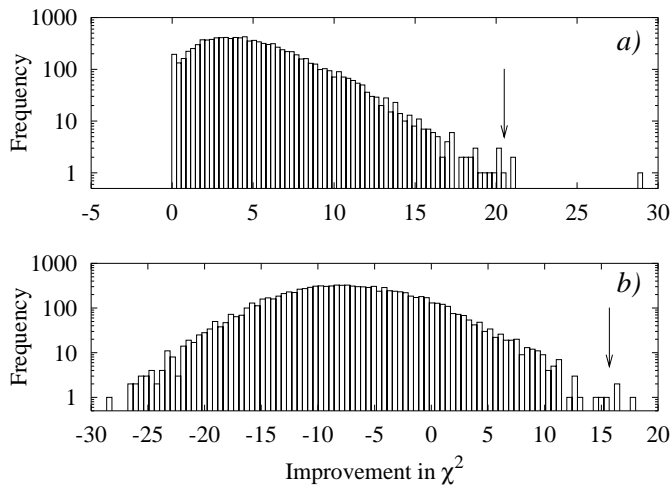}
\caption{Histograms showing the improvements in the fit statistic 
($\chi^{2}$) obtained when
fitting 10000 simulated power-law spectra with; (a) a model consisting of a
power-law plus three emission lines (with the line energies and fluxes
both allowed to vary) and (b) the thermal (\textsc{vmekal}) emission
model. In both cases, only three out of 10000 randomly
generated trial spectra yield an improvement in $\chi^{2}$
equivalent to that found in the actual observation; note the 
$\Delta\chi^{2}$ for the real GRB 011211 data is marked by an arrow. 
The simulations show that the probability of generating 
the soft X-ray line emission through random Poisson
noise is only 0.03\%.}
\label{MC_hists}
\end{figure}

In a recent paper, Rutledge \& Sako (2002) note that the
prior identification of the three strongest emission line 
features with \ion{Si}{xiv}, \ion{S}{xvi} and \ion{Ar}{xviii} represents an {\it a posteriori}
assumption about the data, when trying to estimate the significance of
the line features. However we note (and see Fig.~\ref{mekal}) that
line emission features from these 3 abundant H-like species 
will dominate the emission from a hot (several keV) 
collisionally ionised plasma, or a highly photoionised plasma, over
the 1-4 keV (rest frame) energy range.  
Nonetheless we relax the assumption
that the lines originate from these species, and test the improvement in
fit obtained when adding lines, {\it at any energy from 0.2-10 keV
in the observed spectrum}, to 
the 10000 fake power-law spectra (i.e. a blind spectral search). 
Initially we just add one line to
the fake spectra (whilst varying both the energy and flux of the line), 
and find that an improvement of $\Delta\chi^{2}=8$, 
equivalent to the addition of the \ion{Si}{xiv} line in the real dataset, is
obtained in only 6 out of 1000 cases. Thus the \ion{Si}{xiv} line is detected at
99.4\% confidence. Similarly we find that the \ion{S}{xvi} and \ion{Ar}{xviii} lines
are detected at 98.8\% and 96\% confidence respectively, in agreement
with the values computed by the F-test. We also
consider the significance of the 3 line set as a whole, by fitting 3
Gaussian lines (again at arbitary energies) to the simulated datasets. 
The results (see Fig.~\ref{MC_hists}a) show that only 3 in 10000 trial 
spectra gave an equivalent (or better) 
improvement in the $\chi^{2}$ statistic compared
to the real data (where $\Delta\chi^{2}=20.5$ for the addition of all 3
lines). Thus we can conclude that the addition of 3 Gaussian lines in
the soft X-ray spectrum of GRB 011211 is significant at $>99.9$\%
confidence, whilst the individual lines are detected at 99.4\%,
98.6\% and 96\% confidence (for Si, S and Ar) respectively. 

A more physically realistic model, 
consisting of emission from a hot, thermal plasma (i.e. the 
\textsc{vmekal} model), was also fitted to the 10000 randomly generated
power-law spectra. The temperature, luminosity, redshift 
and relative abundance of the
plasma were allowed to vary in the fits.
Fig.~\ref{MC_hists} (panel b) shows a histogram of the change in
$\chi^{2}$ obtained when comparing a power-law fit to a thermal fit 
for the simulated spectra;
note here a positive change in $\chi^{2}$ corresponds to an improvement
in fit statistic when comparing the \textsc{vmekal} model to a
power-law. {\it In the majority (90\%) of cases, fitting a thermal model
to the randomly generated power-law spectra 
results in a worse fit.} However the real
data for GRB 011211 indicate that an improvement in fit statistic of
$\Delta\chi^{2}=15.7$ is obtained using the thermal model compared 
to the pure power-law. 
Indeed an equivalent improvement in $\Delta\chi^{2}$ 
is only found in 3 out of 10000 simulated spectra; hence 
the inclusion of the
thermal line emission component is formally required at 
$>99.9$\% confidence from the simulations.

\subsection{Other Extraction Effects}

We have also investigated whether varying the size of the 
source selection region has any effect 
on the derived spectrum and the detection of the soft X-ray lines. 
In order to test this, we extracted spectra from circular regions of radii 
10\arcs, 20\arcs\, 30\arcs\ in addition to the previous spectrum that was 
extracted with a 40\arcs\ radius. We find the detection of the redshifted 
soft X-ray line set is significant in each case; at $>$99\% confidence
(via an F-test) 
for spectra extracted with 20\arcs\ and 30\arcs\ radii, whilst the detection 
is less significant (at 98\% confidence) for a 
10\arcs\ radius 
(as substantially fewer source counts are included in the spectrum). 
However, after accounting for the telescope point spread function, 
the derived line flux for 
the \ion{Si}{xiv} line was found to be constant in each case 
(at $\sim1.0\times10^{-5}$~photons~cm$^{-2}$~s$^{-1}$), 
in contrast to the findings of Borozdin \& Trudolyubov (2002) who claim the 
\ion{Si}{xiv} line flux in their spectra decreases with the extraction radius. 

\begin{figure}
\begin{center}
\includegraphics[height=\columnwidth,angle=-90,clip=]{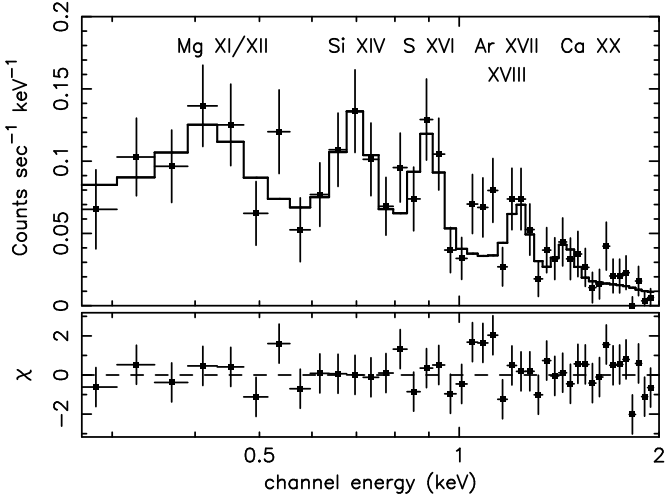}
\end{center}
\caption{EPIC-pn spectrum of GRB 011211, binned into equal energy
bins, 40 eV in size. The 5 redshift emission lines are clearly
detected, and a sixth line at 1.1 keV, due to He-like Ar (\ion{Ar}{xvii}) may also be
present (at 98\% confidence). The likely identifications of the lines
are marked at the top of the plot.}
\label{equalbin_spec}
\end{figure}

The effect of spectral binning was also investigated. Rather than binning the 
spectrum to a set number of photons per bin, the spectra were re-binned so 
that each bin represented a constant energy interval, of 40 eV in size. 
This is then well matched to the resolution of EPIC-pn detector below 
2~keV (where the FWHM spectral resolution is $\sim80$~eV). Data above 
2 keV were ignored due to poor photon statistics (less than 10 source
counts per bin). The resultant spectrum 
is shown in Fig.~\ref{equalbin_spec}. The spectrum has been fitted with the 5 redshifted 
lines obtained previously, the fit statistic again improves significantly 
(from $\chi^{2}/dof=60.5/42$ to 33.4/32, i.e. at 99\% confidence) 
upon the addition of the line set. Interestingly, a residual 
near 1.1 keV is present in the data, the addition of a sixth 
emission line, at $1.09\pm0.02$~keV, is significant at 98\% confidence 
($\chi^{2}/d.o.f=25.9/30$). Assuming a redshift of $z=1.9$, this line 
could correspond to He-like Ar (\ion{Ar}{xvii}), which is predicted in the 
thermal model (Fig.~\ref{mekal}). The net redshift obtained for the line set is 
then $z=1.86\pm0.04$, consistent with the previous determination
(section 3.1).

\section{Temporal variability of the Line Emission in GRB 011211}

The observations presented in Reeves \et (2002) showed that the line 
emission in GRB 011211 decayed with time faster than the continuum, 
rendering the lines undetectable after 10 ks of observation. To investigate 
this further, we initially split the first 12 ksec of observations into 
4$\times$3~ksec segments. Although it is difficult to constrain the 
significance of the {\it individual} lines over such a short time period, 
nonetheless we compare the power-law model with the 
thermal (\textsc{vmekal}) model 
and the model consisting of a power-law plus redshifted emission lines. 
The results of the fits are shown in table 3. 


\begin{table*}
\centering
\caption{Fits to GRB 011211, over the first 12 ksec of
observation. $^{a}$ Time measured from start of observation. $^{b}$
Power-law model with Galactic absorption. $^{c}$ Power-law model with
free neutral absorption (measured in observed frame). $^{d}$ Units of 
column density $10^{21}$~cm$^{-2}$. $^{e}$ Fit using variable
abundance thermal plasma model \textsc{vmekal}. $^{f}$ Indicates
parameter is fixed in fit. $^{g}$ Redshift determined from
model. $^{h}$ Power-law model including 5 redshifted emission lines
from Mg, Si, S, Ar and Ca.}

\begin{tabular}{lcccccc}
\hline\hline                 

Model & & 0-3~ksec$^{a}$ & 3-6~ksec$^{a}$ & 6-9~ksec$^{a}$ &
9-12~ksec$^{a}$ \\

\hline

PL only$^{b}$ & $\Gamma$ & $2.15\pm0.16$ & $2.32\pm0.21$ & $2.14\pm0.06$ &
$2.42\pm0.20$ \\

& $\chi^{2}/d.o.f$ & 51.5/28 & 23.9/16 & 26.8/24 & 18.7/26 \\

\hline

PL*wabs$^{c}$ & $\Gamma$ & $3.6\pm0.7$ & $4.3\pm1.1$ & $2.9\pm0.5$ &
$2.42\pm0.20$ \\

& $N_{H}$$^{d}$ & $3.0\pm1.1$ & $3.6\pm1.5$ & $1.6\pm0.8$ & $<2.5$ \\

& $\chi^{2}/d.o.f$ & 39.1/27 & 12.1/15 & 23.6/23 & 16.9/26 \\ 

\hline

Thermal$^{e}$ & $kT$ & $4.1\pm1.2$ & $3.3\pm0.8$ & $3.1\pm0.6$ &
$2.3\pm0.5$ \\

& z$^{g}$ & $1.89\pm0.09$ & $1.89\pm0.12$ & $1.9^{f}$ & $1.9^{f}$ \\

& $\chi^{2}/d.o.f$ & 31.9/26 & 14.0/14 & 20.6/22 & 14.1/24 \\ 

\hline

PL + z-lines$^{h}$ & $\Gamma$ & $3.0\pm1.1$ & $2.4\pm0.5$ & -- & -- \\

& z$^{g}$ & $1.90\pm0.08$ & $1.86\pm0.07$ & -- & -- \\

& $\chi^{2}/d.o.f$ & 21.9/22 & 11.4/10 & -- & -- \\ 

\hline

\end{tabular}
\end{table*}

It can be seen that the two spectra from the initial 6 ksec of observation 
are poorly fit by a simple power-law spectrum, 
whilst both the redshifted line and 
thermal emission models are a significantly better fit. What appears 
interesting is that the early afterglow spectrum 
is significantly curved. The spectrum 
during the first 3 ksec can be fit with a steep power-law 
(with $\Gamma=3.6\pm0.7$) and an observed frame 
column density 
of $N_{H}=3.0\pm1.1\times10^{21}$~cm$^{-2}$. This indicates that 
the early afterglow spectrum is 
dominated by an intrinsically curved (thermal) component. 
In contrast, acceptable fits to the last two spectra (between 6 and 12 ksec) 
can be obtained by a pure power-law model, neither the soft X-ray line 
emission, nor the 
addition of a thermal emission component being formally required in these 
spectra. 

In order to quantify this, we have calculated the ratio of thermal to
power-law emission during the first and last 10 ks of
observation. During the first 10 ksec, the thermal component has an
average luminosity of $2.0\times10^{45}$~erg~s$^{-1}$, whilst the
power-law component has a luminosity of $6.8\times10^{45}$~erg~s$^{-1}$
(and note that both components are statistically 
required in the spectral fit at $>99$\% confidence). In
the last 10 ksec the power-law component has a luminosity
$4.85\times10^{45}$~erg~s$^{-1}$, whilst the thermal emission is
no longer required ($L<1.07\times10^{45}$~erg~s$^{-1}$). 

We have also parameterised the strength of the line
emission by measuring the equivalent width (EW) of the \ion{Si}{xiv} line,
relative to a pure power-law continuum. 
During the first 5 ksec, the \ion{Si}{xiv} line is detected with an (observed frame) 
EW of $127\pm42$~eV. A line is also marginally detected (at 90\%
confidence) from 5-10 ksec after the start of the observation, with a
smaller EW of $62\pm30$~eV. Beyond 10 ks, the line is not detected, we
can place a tight upper-limit of $<25$~eV on the EW of the line from
the last 17 ks of observation. Thus clearly the soft X-ray line
emission in GRB 011211 decays with time, and is not detected after 10
ks from the start of the \xmm\ observation. 

\section{Discussion and Conclusions}

Both the F-test and Monte-Carlo simulations have showed 
that the soft X-ray line emission in GRB 011211 is 
detected at $>99\%$ confidence, 
We conclude that the 
detection of the soft X-ray features in GRB 01211 is robust, and 
is not affected by the detector background, calibration, 
spectral binning or the particular spectral model that is assumed.  
It is also apparent that the lines decay rapidly, no line emission is 
formally detected after 10~ks from the start of the \xmm\ 
observation. 

The early afterglow data (during the first 10 ksec of 
observation) appears to be characterised by thermal X-ray emission (from 
optically thin gas), as evident by the curvature present in 
the in the early spectrum. 
An alternative spectral model, where the 
afterglow line emission results from X-ray reflection off optically thick 
material appears to be ruled out (e.g. Ballantyne \& Ramirez-Ruiz
2001), although 
we note that in current reflection models, the emission from S, Ar and Ca 
is not computed. Future models will hopefully provide a 
better test of the reflection scenario.

On the assumption that the lines result from thermal emission one can 
estimate the mass of the ejected material. Firstly the emission measure 
($E_{M}$) is computed from the observed luminosity of the thermal emission
component. Here $n^{2}V=E_{M}\sim10^{69}$cm$^{-3}$ 
(where n is the electron density in cm$^{-3}$ and V is the volume), 
whilst the plasma will emit X-rays over a characteristic cooling time
(governed by the electron density) 
of $t_{cool}=1.4\times10^{15}n^{-1}$~secs, for gas of temperature $kT=4$~keV. 
Thus from the emission measure, the total integrated 
output of the thermal emission is
governed by $n^{2}V=E_{M}t_{line}/t_{cool}$, where $t_{line}$ is the 
observed lifetime of the line emission.
Substituting for the cooling time, and as $nV=M/m_{p}$ (where 
M is the total mass of the emitting material and $m_{p}$ is the proton mass), 
then one can derive the total mass of the ejecta:-

\begin{center}

$M = 2.4\times 10^{30} t_{line}$~g \\
or $M = 1.2 \times 10^{-3} t_{line}~M_{\odot}$
 
\end{center}

The minimum lifetime of the line emission (measured from the start of the 
\xmm\ observation) is $10^{4}/(1+z)$~secs, whilst the maximum 
lifetime occurs if the line emission starts at the time of the burst, 
which is then $t_{line}=5.5\times10^{4}/(1+z)$~secs. 
Thus the minimum and maximum mass of 
the outflowing material, responsible for the blue-shifted line emission, are 
then 4 and 20 solar masses respectively. This indicates that the mass of 
the stellar progenitor is likely to be $>20M_{\odot}$  
even for the conservative case where the mass of the {\it ejecta} 
is $4M_{\odot}$ (e.g. Woosley \& Weaver 1995).

The duration of the line features in GRB 011211, observed 11 hours 
after the initial burst, implies a distance of several~$\times~10^{15}$~cm 
for the 
line emitting matter (Reeves \et 2002). When combined with the 
velocity of the material ($\sim0.1c$), 
this implies that the matter was 
ejected from days to weeks before the burst itself occurred. 
This time delay appears too short in the context of the ``supranova'' 
model (Vietri \& Stella 1998), where a time-delay of several months 
to years can be predicted, whilst the ``hypernova'' class of models 
(Woosley 1993) involve the near-simultaneous occurrence of a supernova 
and the gamma ray burst. Interestingly one model has been 
proposed involving the coalescence of a close binary system, resulting from 
a collapse of a massive stellar progenitor (Davies \et 2002). This 
could naturally account for the time delay between the putative 
supernova and the gamma ray burst, with a delay of a few days resulting 
from the orbital decay (through gravitational radiation) of the close 
binary system.

If the line emitting material is located at much smaller 
distances ($R\sim10^{13}$~cm), as in nearby reprocessor models (e.g. 
Rees \& Meszaros 2000) then the need for a significant time delay is 
reduced. Here, the line emission results from re-processing 
(via reflection/photoionisation) of continuum photons emitted 
{\it after} the initial burst. 
However in this scenario, significant continuum emission should be observed  
at the same time as the line emission, which would render any 
soft X-ray line features undetectable by current instrumentation (see 
Lazzati \et 2002). One possibility that has been suggested (Kumar 
\& Narayan 2002) is that the ionising 
photons are scattered back towards the ejected material, by a 
positron-electron shield generated by the initial $\gamma$-rays 
themselves. This model 
removes the need for a supernova-GRB delay, as the duration of the 
line emission is accounted for by the time required for the photons to be 
scattered back towards the ejected matter. 

Including GRB 011211, there have now been 4 detections of GRB afterglows 
with \xmm\ (GRB 001025A and GRB 010220, Watson \et 2002; GRB 020322, 
Ehle \et 2002), whilst the X-ray afterglow for the 
faint burst GRB 020321 could not be localised. 
Interestingly, line emission features are also indicated 
in the X-ray afterglows 
of GRB 001025A and GRB 010220 (Watson \et 2002). Similar to GRB 011211, 
the spectrum of GRB 001025A appears to exhibit a blend of soft X-ray line 
emission, from the medium-z elements such as Mg, Si, S, and Ar. 
In contrast, the spectrum of GRB 010220 shows a strong 
iron-group (i.e. Fe, Co or Ni) line (equivalent width $\sim1$~keV), 
although here the X-ray spectrum is absorbed below 1 keV, rendering
any soft X-ray line emission undetectable. 
Both fits favour an over abundance of Nickel (or 
Cobalt) to iron, which is suggestive of a short time delay (days 
rather than months) between the putative supernova and burst.
Thermal emission models are formally required to fit the soft 
X-ray excess observed in GRB 001025A, whilst the spectrum of 
GRB 010220 can be equally 
well fitted by either thermal or reflection models. Thus soft X-ray 
line emission is required in at least 2 of the \xmm\ afterglow spectra
(GRB 011211 and GRB 001025A), the exception being the recent burst 
GRB 020322, which  appears to exhibit a featureless (but absorbed) 
power-law spectrum (Reeves \et 2002b, in preparation). 

It seems possible that soft X-ray line emission features are relatively 
common in X-ray afterglow spectra, whilst thermal emission models can fit 
3 out of the 4 present \xmm\ spectra. 
The previous lack of detection of soft X-ray 
features with Chandra and \sax\  
can be accounted for by the 
lower effective area of the 
ACIS-S and LECS detectors respectively in the soft X-ray band.
Indeed \xmm\ is the first X-ray mission with the sensitivity required 
to detect any putative soft X-ray features in GRBs at energies below 2 keV.
It is important that \xmm\ performs follow-up 
observations of bright ($>10^{-13}$~erg~cm$^{-2}$~s$^{-1}$) 
afterglows, within several hours of the burst, to determine the frequency 
the soft X-ray line emission. 
Ultimately, through monitoring the temporal behavior of the 
line emission from minutes to many hours after a burst, 
the X-ray telescope on-board NASA's forthcoming {\it Swift} mission 
will be able to trace 
the distribution of matter around the burst explosion and solve the 
ambiguities that arise when discriminating between the current 
GRB models.

\acknowledgements

This paper is based on observations obtained with XMM-Newton, an ESA
science mission with instruments and contributions directly funded by
ESA Member States and the USA (NASA). 
James Reeves would like to thank the Leverhulme Trust
organisation for their support, in the form of a Research Fellowship.

\label{lastpage}
\end{document}